\documentclass[twocolumn,aps,prb,showpacs,superscriptaddress]{revtex4}
\usepackage{graphicx}\usepackage{amsmath}\usepackage{amsfonts}\usepackage{amssymb}

\begin{document}
\author{Arne Brataas}\affiliation{Department of Physics, Norwegian University of Science and Technology,N-7491 Trondheim, Norway}
\author{Yaroslav Tserkovnyak}\affiliation{Lyman Laboratory of Physics, Harvard University, Cambridge, Massachusetts02138, USA}
\author{Gerrit E. W. Bauer}\affiliation{Kavli Institute of NanoScience, Delft University of Technology, Lorentzweg1, 2628 CJ Delft, The Netherlands}
\title{Current-induced macrospin \textit{vs.} spin-wave excitations in spin valves}

\begin{abstract}The mode dependence of current-induced magnetic excitations in spin valves is studied theoretically. The torque exerted on the magnetization by transverse spin currents as well as the Gilbert damping constant are found to depend strongly on the wave length of the excitation (spin wave). Analytic expressions are presented for the critical currents that excite a selected spin wave. The onset of macrospin (zero wavelength) \textit{vs.} finite wavelength instabilities depends on the device parameters and the current direction, in agreement with recent experimental findings. \end{abstract}

\date{\today}
\maketitle

\section{Introduction}
Less than a decade ago Berger \cite{Berger:prb96} and Slonczewski \cite{Slonczewski:mmm96} argued that an electric current sent through multilayers of normal metals (N) and ferromagnets (F) can excite the ferromagnetic order parameter and even reverse the magnetization. The theoretical predictions have been confirmed by many experiments on F$\mid$N$\mid$F nanostructured spin valves.\cite{Tsoi:prl98,Ozyilmaz:prl03} The physics of collective ferromagnetic excitations driven by out-of-equilibrium quasi-particles is complex and fascinating. In magnetic memories current-induced magnetization switching might turn out to be superior to its magnetic field driven counterpart. 

Current-induced magnetic excitations are driven by the spin transfer torque acting on the magnetic order parameter when a spin current polarized normal to the magnetization is absorbed by the ferromagnet.\cite{Berger:prb96, Slonczewski:mmm96,Brataas:prl00,Waintal:prb00} The transverse spin current extinction is a quantum mechanical dephasing effect between electrons at the Fermi energy with different precession lengths. In Co, Ni and Fe this happens on an atomistic length scale.\cite{Brataas:prl00,Waintal:prb00} By conservation of angular momentum the absorbed spin current acts as a torque on the ferromagnetic condensate. At a critical spin current, this torque becomes strong enough to set the magnetization into motion, possibly leading to a complete reversal of the magnetization direction.

One discussion that started with the prediction of the current-induced magnetization dynamics remains to be settled. Berger defines the critical current at the onset of spin wave excitations, whereas Slonczewski considers a rigid coherent rotation of the whole magnet (``macrospin model). In the latter scenario, the critical current corresponds to a torque that just overcomes the Gilbert damping. In our view, the differences in these pictures are to some extent semantic, since the macrospin model is identical to the lowest energy spin wave. The physical question addressed here is the wavelength of the spin wave that is most easily excited. We find that there is no universal answer and that the preferential excitation mode is a complicated function of device parameters and current direction. Nevertheless, our theory agrees well with experiments that observe both type of excitations.\cite{Ozyilmaz:04}

Most theories of spin transfer torques and critical currents are based on macrospin precessions in spin valves.\cite{Slonczewski:mmm96,Waintal:prb00} Indeed, sufficiently small magnetic clusters support a single domain magnetization and the magnetic field induced magnetization reversal is well described by a coherent rotation according to the Stoner-Wohlfarth model.\cite{Wernsdorfer} From a theoretical point of view, the macrospin models validity is essential for understanding the non-linear physics underlying the entire magnetization dynamics by dynamical systems theory and the probabilistic treatment in the ``presence'' of noise.\cite{Apalkov} In a single ferromagnetic film sandwiched by normal layers, the torque on the macrospin domain necessarily vanishes. However, biased N$\mid$F$\mid$N structures are unstable with respect to spin waves with shorter wavelengths.\cite{Polianski:prl04,Stiles:prb04} In spin valves, we may therefore expect a competition between macrospin and shorter wave length spin waves excitations. Experiments on spin valves \cite{Ozyilmaz:04} have indeed been interpreted in terms of both types of excitations, depending on the current direction (and the spin accumulation pattern). It is our purpose to understand and model these data in order to assess the dependence of the excitation modes on device parameters. Another motivation to study the competition between different excitation modes is the need to find criteria for the breakdown of simple models for the magnetization that provide a guide for the necessity of full-fledged micromagnetic calculations.\cite{Lee:natmat04}

Our work extends Refs.~\onlinecite{Polianski:prl04,Stiles:prb04} on single layers to spin valves. We derive analytical expressions for bias-driven spin transfer torques, enhanced Gilbert damping constants \cite{Tserkovnyak:prl02} and critical currents for magnetic excitations as a function of wave vector. We predict a rich ``phase diagram'' in the current \textit{vs.} magnetic field plane.

Our paper is organized as follows. Section~\ref{s:model} introduces our model for the spin valve pillar. We also present the assumptions used in computing the charge current, the spin current, the spin transfer torques and the enhanced Gilbert damping governed by spin pumping. Furthermore, a schematic description of the necessary ingredients for the calculations is outlined. Section~\ref{s:current} presents the computational details for readers interested in a deeper knowledge on how our results have been obtained. Our results are presented and discussed in section~\ref{s:results}. In that section, we also compare our results with the measurements by the NYU/IBM collaboration and find a semi-quantitative agreement. We conclude our paper in section ~\ref{s:conclusions}.

\section{Model and assumptions}\label{s:model}

\begin{figure}[tbp]\includegraphics[width=\linewidth,angle=0,clip=true,width=8cm]{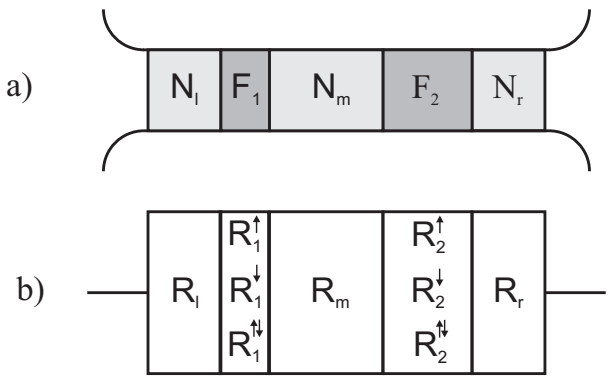} 
\caption{Spin-valve a)with circuit resistance elements sketched in b).}
\label{spinvalve}
\end{figure}

We consider N$\mid$F$\mid$N$\mid$F$\mid$N spin-valve pillars in Fig.\  \ref{spinvalve}a in the semiclassical transport regime where magneto-electronic circuit theory applies.\cite{Brataas:prl00}  Disregarding spin flips in F is allowed for sufficiently thin F films and considerably simplifies the calculations. We also consider normal metals thinner than the spin-diffusion length. A possibly spatially dependent pillar cross section, not discussed below any further, can be treated by simply scaling the resistance parameters. 

We compute the spin transfer torques, the enhanced Gilbert damping and the critical current for magnetic excitations in the following way. The charge and spin accumulation pattern and the charge and spin current through the system is computed by treating the N$\mid$F$\mid$N$\mid$F$\mid$N spin valve layer by layer. Our calculations includes scattering due to bulk impurities within each layer {\it and} scattering due to the band-structure mismatch between adjacent metals, \textit{e.g.}\ interface resistances. Both bulk impurity scattering and interface mismatch are important in \textit{e.g.}\ understanding the giant magnetoresistance effect in magnetic multilayers. Bulk impurity scattering is taken into account by solving the diffusion equations in the bulk of each material. Spin and charge current through four interfaces between the normal metals and the ferromagnets are determined by the interface scattering matrix between the adjacent materials which gives rise to the interface resistances that we take into account. 

We will show below that our results for the spin transfer torques, the enhanced Gilbert damping and the critical current for magnetic excitations can be expressed in terms of the resistance parameters of the electronic circuit in Fig.\ \ref{spinvalve}b. Let us first introduce the resistances that are important for the calculations and the results: Starting from the left in Fig.\ \ref{spinvalve}b we introduce a bulk resistance per unit area for the bulk of the left normal metal, $R_l=\rho_{N_l} t_{N_l}$, where $\rho_{N_l}$ is the resistivity of the normal metal and $t_{N_l}$ is the thickness of the normal metal layer in the transport direction. Similarly, starting from the right, we introduce the bulk resistance per unit area for the bulk of the right normal metal, $R_r=\rho_{N_r} t_{N_r}$ and the bulk resistance per unit area for the bulk of the middle normal metal, $R_m=\rho_{N_m} t_{N_m}$. The interesting physics arises due to the spin-dependent scattering within the bulk of the ferromagnetic layers \textit{and} at the interfaces between the normal metals and the ferromagnetic layers. The spin-dependent resistances $R_{i}^s$ ($i=1,2$, $s=\uparrow, \downarrow$) in the ferromagnetic elements consist of interface and bulk contributions: $R_{i}^{s}=2R_{F_{i}/N}^{s}+\varrho _{F_{i}}^{s}t_{F_{i}}$, where $R_{F_{i}/N}^{s}$ is the spin-dependent resistance per unit area of a single F$\mid$N interface assumed identical on both sides of film $i$. $\varrho _{F_{i}}^{s}$ and $t_{F_{i}}$ are its spin-dependent resistivity and thickness.  When the magnetizations are collinear the spin dependent resistances in the ferromagnets with the resistances in the normal metals are sufficient to describe the transport of charge and spin in the systems. When the magnetizations are non-collinear, the flow of spins directed perpendicular to the magnetization in the ferromagnets must also be quantified. The mixing conductance $G_i^{\uparrow \downarrow}$ determines the absorption of spins transverse to the magnetization direction and consequently the spin transfer torque.\cite{Brataas:prl00}  We find it convenient to introduce a mixing resistance $R_i^{\uparrow \downarrow }=1/(2G_i^{\uparrow \downarrow})$ so that all our results can be presented in terms of resistances. 

The charge and spin currents are not only induced by the applied bias, but also by spin pumping due to a moving magnetization that opposes the magnetization dynamics.\cite{Tserkovnyak:prl02} Effectively, when a layer magnetization precesses, it acts as a source of a spin current transverse to the magnetization direction and proportional to the precessional frequency, which should be included in the self-consistent computation for the charge and spin currents in magnetoelectronic circuits. When spin-waves are excited in one of the layers, the sources of spins are non-uniform. This leads to a wave-vector dependence of the enhanced Gilbert damping. These contributions can be computed by using the same equivalent circuit, Fig.\ \ref{spinvalve}b, as for the bias voltage induced charge and spin current and also depend on the same resistance parameters of the circuit. The wave-vector dependent enhanced Gilbert damping turns out to be governed by the magnitude of the spin-pumping current out of the ferromagnet and the resistance between the ferromagnet and either the voltage reservoirs or other spin sinks for the emitted spin current.

We wish to compute the critical currents for the magnetic instabilities in driven spin-valve pillars by linearizing the equations that govern and the magnetization dynamics close to the equilibrium configurations. In this way, we can find the critical currents for magnetic excitations, but we cannot unambiguously predict what happens when the current is increased beyond the critical current. To the latter end, one needs to go beyond the linear instability regime and study the full micromagnetic behavior of each individual ferromagnet as well as the transport of spin and charge self-consistently. This is beyond the scope of the present work. We would like to emphasize that even in the linear regime the dynamics of the ferromagnets and the transport of spin and charge have to be treated self-consistently. It is the self-consistent feedback of the spin-flow from a non-uniform domain back into the ferromagnet that destabilizes the monodomain configuration in a the single layer ferromagnet.\cite{Polianski:prl04}

We are mostly interested in spin-valves that have a relatively large in-plane magnetic field along the magnetization direction as \textit{e.g.} studied by the NYU/IBM collaboration. Therefore, we proceed from equilibrium configurations of monodomain magnetizations in parallel or anti-parallel to each other. Out of equilibrium, spin waves can be excited transverse or parallel to the transport direction. For sufficiently thin ferromagnets, the critical current for the transverse excitations are lower than the critical current for longitudinal excitations.\cite{Polianski:prl04,Stiles:prb04} Consequently, we consider small transverse instabilities of the magnetization direction in Fourier space, $\delta \mathbf{m}_{1}^{\perp }(\mathbf{q})$ and $\delta \mathbf{m}_{2}^{\perp }(\mathbf{q})$, where $\mathbf{q}$ is a two dimensional vector in Fourier space transverse to the transport direction and compute the perturbed charge and spin current through the system. The perturbed spin currents induce spin transfer torques on the ferromagnetic layers. Also, a precessing ferromagnet emits spin current into the adjacent materials than can enhance the Gilbert damping. We compute whether the magnetic instabilities are stable or unstable and find the critical current for magnetic excitations. Longitudinal magnetic excitations along the transport direction (as studied \textit{e.g.} in Ref.\ \onlinecite{Stiles:prb04}) are thus disregarded. We will comment on this thin-layer assumption when applying our theory to the NYU/IBM experiments in Sec. \ref{s:results}.

\section{Calculation of current}\label{s:current}
We introduce charge $V^{(c)}$ and spin ${\bf V}^{(s)}$ potentials in the normal metals close to the N$\mid$F interfaces. These potentials depend on the two-dimensional coordinate $\text{\boldmath$\rho$}=(x,y)$ along the interface transverse to the transport direction $z$. In our notation, $j_{1l}$ ($j_{1m}$) denotes the charge current density incident from the left (middle) normal metal and going through ferromagnet $1$ with magnetization along unit vector $\mathbf{m}_1$, and $j_{2m}$ ($j_{2r}$) denotes similarly the charge current density incident from the middle (right) normal metal and going through ferromagnet 2 with magnetization along unit vector $\mathbf{m}_2$. The directions of the magnetizations also depend on the transverse coordinate \text{\boldmath$\rho$} allowing spin-wave excitations in the transverse direction, $\mathbf{m}_1=\mathbf{m}_1(\text{\boldmath$\rho$})$ and $\mathbf{m}_2=\mathbf{m}_2(\text{\boldmath$\rho$})$. The charge current along the N$\mid$F interface thus depends on the transverse coordinate \text{\boldmath$\rho$}. The charge currents on the two sides of the first and second ferromagnet-normal metal interface are
\begin{align}
j_{1l}^{(c)} & =( G_{1}^{\uparrow }+G_{1}^{\downarrow }) \Delta V_{1}^{(c)}+( G_{1}^{\uparrow }-G_{1}^{\downarrow }) \Delta {\bf V}_{1}^{(s)}\cdot {\mathbf{m}}_{1}\,, \label{jc_1l} \\
j_{1m}^{(c)} & =j_{1l}^{(c)}\,, \label{jc_1m} \\
j_{2m}^{(c)} & =( G_{2}^{\uparrow }+G_{2}^{\downarrow }) \Delta V_{2}^{(c)}+( G_{2}^{\uparrow }-G_{2}^{\downarrow }) \Delta {\bf V}_{2}^{(s)}\cdot {\mathbf{m}}_{2}\,, \label{jc_2m} \\
j_{2r}^{(c)} & = j_{2m}^{(c)} \,,\label{jc_2r}
\end{align}
where $\Delta V_{1}^{(c)}=V_{1m}^{(c)}-V_{1l}^{(c)}$  and  $\Delta V_{2}^{(c)}=V_{2r}^{(c)}-V_{2m}^{(c)}$ are the charge voltage drop over the ferromagnets 1 and 2. The spin voltage drop $\Delta {\bf V}_{1}^{(s)}={\bf V}_{1m}^{(s)}-{\bf V}_{1l}^{(s)}$ and $\Delta {\bf V}_{2}^{(s)}={\bf V}_{2r}^{(s)}-{\bf V}_{2m}^{(s)}$ are defined analogously. Charge current is conserved on traversing the ferromagnet, which is reflected in the conditions $j_{1l}^{(c)}=j_{1m}^{(c)}$ and $j_{2m}^{(c)}=j_{2r}^{(c)}$. The spin-dependent conductances $G_{i}^s$ ($i=1,2$ and $s=\uparrow,\downarrow$) consist of interface and bulk contributions: $1/G_{i}^{s}=R_{i}^s=2R_{F_{i}/N}^{s}+\varrho _{F_{i}}^{s}t_{F_{i}}$, where $R_{F_{i}/N}^{s}$ is the spin-dependent resistance per unit area of a single F$\mid$N interface assumed identical on both sides of film $i$. $\varrho _{F_{i}}^{s}$ and $t_{F_{i}}$ are its spin dependent resistivity and thickness. 
Similarly to the charge current, we express the spin current density incident from the left (middle) normal metal and going into ferromagnet 1 as ${\mathbf j}_{1l}$ ($\mathbf{j}_{1m})$ and the spin current density incident from the middle (right) normal metal and going into ferromagnet 2 as $\mathbf{j}_{2m}$ ($\mathbf{j}_{2r}$). The spin-current density is expressed in terms of parallel ($\parallel$) and transverse components ($\perp$),
\begin{align}{\mathbf j}_{1l}^{(s)} & =j_{1l}^{(s \parallel )}{\mathbf m}_{1}+{\mathbf j}_{1l}^{(s\perp )} \, , \label{js_1l} \\
{\mathbf j}_{1m}^{(s)} & =j_{1m}^{(s \parallel )}{\mathbf m}_{1}+{\mathbf j}_{1m}^{(s\perp )} \, ,  \label{js_1m} \\
{\mathbf j}_{2m}^{(s)} & =j_{2m}^{(s \parallel )}{\mathbf m}_{2}+{\mathbf j}_{2m}^{(s\perp )} \, , \label{js_2m} \\
{\mathbf j}_{2r}^{(s)} & =j_{2r}^{(s \parallel )}{\mathbf m}_{2}+{\mathbf j}_{2r}^{(s\perp )} \, . \label{js_2r}
\end{align}
The parallel components are
\begin{align}j_{1l}^{(s\parallel )} & = ( G_{1}^{\uparrow }-G_{1}^{\downarrow}) \Delta V_{1}^{(c)} + ( G_{1}^{\uparrow }+G_{1}^{\downarrow}) \Delta {\bf V}_{1}^{(s)}\cdot {\mathbf m}_{1}\,, \\j_{1m}^{(s\parallel )} & = j_{1l}^{(s\parallel)} \,,\\j_{2m}^{(s\parallel )} & = ( G_{2}^{\uparrow }-G_{2}^{\downarrow}) \Delta V_{2}^{(c)} + ( G_{2}^{\uparrow }+G_{2}^{\downarrow}) \Delta {\bf V}_{2}^{(s)}\cdot {\mathbf m}_{2}\,,\\j_{2r}^{(s\parallel )} & = j_{2m}^{(s\parallel)} \,. \label{js_par}\end{align}
The component of the spin-current parallel to the magnetization direction is conserved on traversing the ferromagnet which is reflected in the conditions $j_{1l}^{(s\parallel)}=j_{1m}^{(s\parallel)}$ and $j_{2m}^{(s\parallel)}=j_{2r}^{(s\parallel)}$. The transverse components of the spin-current densities are not conserved on traversing the interface due to the absorption of transverse spin flow:\cite{Brataas:prl00}
\begin{align}
{\mathbf j}_{1l}^{(s\perp )} & =-2G_{1}^{\uparrow \downarrow }{\mathbf m}_{1}\times {\bf V}_{1l}^{(s)}\times {\mathbf m}_{1}\,, \label{js_perp_1l}\\
{\mathbf j}_{1m}^{(s\perp )} & =2G_{1}^{\uparrow \downarrow }{\mathbf m}_{1}\times {\bf V}_{1m}^{(s)}\times {\mathbf m}_{1}\,, \label{js_perp_1m} \\
{\mathbf j}_{2m}^{(s\perp )} & =-2G_{2}^{\uparrow \downarrow }{\mathbf m}_{2}\times {\bf V}_{2m}^{(s)}\times {\mathbf m}_{2}\,, \label{js_perp_2m}\\
{\mathbf j}_{2r}^{(s\perp )} & =2G_{2}^{\uparrow \downarrow }{\mathbf m}_{2}\times {\bf V}_{2r}^{(s)}\times {\mathbf m}_{2} \,, \label{js_perp_2r}
\end{align}
where it is assumed that the mixing conductance is the same for the interface between the left (middle) normal metal and ferromagnet 1 (2) as for the interface between the middle (right) normal metal and ferromagnet 1 (2), and the small imaginary part of the mixing conductance $G_{i}^{\uparrow \downarrow }$ has been disregarded.\cite{Brataas:prl00} The spin-torque densities on ferromagnets 1 and 2 are determined by the absorption of the transverse spin currents:
\begin{eqnarray}\text{\boldmath$\tau_1$} & = & \mathbf{j}_{1l}^{(s\perp)}-\mathbf{j}_{1m}^{(s\perp)} \, ,\\\text{\boldmath$\tau_2$} & = & \mathbf{j}_{2m}^{(s\perp)}-\mathbf{j}_{2r}^{(s\perp)} \, .
\end{eqnarray}
>From the expressions for the spin current densities (\ref{js_1l}), (\ref{js_1m}), (\ref{js_2m}) and (\ref{js_2r}) and the flow of spins transverse to the magnetizations (\ref{js_perp_1l}), (\ref{js_perp_1m}), (\ref{js_perp_2m}) and (\ref{js_perp_2r}), we can express the spin transfer torque densities in terms of the spin accumulations in the normal metals:
\begin{align}
\text{\boldmath$\tau_1$} & =  - 2 G_1^{\uparrow \downarrow} {\mathbf m}_1 \times \left[({\bf V}_{1l}^{(s)} + {\bf V}_{1m}^{(s)}) \times {\mathbf m}_1 \right] \, , \\
\text{\boldmath$\tau_2$} & =  - 2 G_2^{\uparrow \downarrow} {\mathbf m}_2 \times \left[({\bf V}_{2m}^{(s)} + {\bf V}_{2r}^{(s)}) \times {\mathbf m}_2 \right] \, .
\end{align}

We consider normal metals thinner than the spin-diffusion length, but spin-flip may not be neglected. Eqs.\ (\ref{jc_1l}), (\ref{jc_1m}), (\ref{jc_2m}), and (\ref{jc_2r}) and (\ref{js_1l}), (\ref{js_1m}), (\ref{js_2m}) and (\ref{js_2r}) should be matched to solutions of the diffusion equations for charges and spins
\begin{equation}\nabla ^{2}V_{\eta}^{(c)}=0\,,\,\,\nabla ^{2}{\mathbf V}_{\eta}^{(s)}= {\mathbf V}_{\eta}^{(s)}/l_{\eta}^{2}   
\label{diff_eq} \, ,
\end{equation}
where $\varrho_{N\eta}$ is the resistivity and $l_\eta$ the spin-diffusion length in each normal metal layer with index $\eta={l,m,r}$. We denote the direction along the spin valve pillar, \textit{e.g.} the transport direction, $z$. The charge and spin currents in the transport direction are 
\begin{equation}
j_\eta^{(c)}(z)=\frac{1}{\varrho_{Ni}} \frac{\partial V_{\eta}^{(c)}}{\partial z}\,,\,\,\mathbf{j}_\eta^{(s)}(z)=\frac{1}{\varrho_{Ni}} \frac{\nabla \mathbf{V}_{\eta}^{(s)}}{\partial z} \,.\label{diff_cur}
\end{equation}
The thicknesses of the left, middle and right normal metals are $L_l$, $L_m$, and $L_r$, respectively. We assume that the left and right normal metals are attached to reservoirs where charges and spins are in local equilibrium. We choose local coordinates, $z_l$, $z_m$, and $z_r$, for each normal metal so that 1) in the left normal metal $z_l=0$ corresponds to the contact between the left side normal metal and the left reservoir and $z_l=L_l$ corresponds to the N$_l$$\mid$F$_1$ interface, 2) in the middle normal metal $z_m=0$ corresponds to the right side of the F$_1$$\mid$N$_m$ interface and $z_m=L_m$ corresponds to the left of the N$_m$$\mid$F$_2$ interface, and 3) in the right normal metal $z_r=0$ corresponds to the right side of the F$_2$$\mid$N$_r$ interface and $z_r=L_r$ corresponds to the boundary between the right normal metal and the reservoir.

Conservation of spin and charge requires continuity of spin and charge flow in the bulk of the normal metals close to the left and to the right of the N$\mid$F interfaces and through the interfaces:
\begin{align}
j_{1l}^{(c)} & = j_l^{(c)}(z_l=L_l) \,,\,\, \mathbf{j}_{1l}^{(s)}  = \mathbf{j}_l^{(c)}(z_l=L_l) \,, \\
j_{1m}^{(c)} &= j_m^{(c)}(z_m=0) \,,\,\, \mathbf{j}_{1m}^{(s)} = \mathbf{j}_m^{(c)}(z_m=0) \,, \\
j_{2m}^{(c)} &= j_m^{(c)}(z_m=L_m) \,,\,\, \mathbf{j}_{2m}^{(s)}= \mathbf{j}_m^{(c)}(z_m=L_m) \,,\\
j_{2r}^{(c)} &= j_r^{(c)}(z_r=0) \,,\,\, \mathbf{j}_{2r}^{(s)}= \mathbf{j}_r^{(c)}(z_r=0) \, .
\end{align}
On the outer, left and right, normal metal layer, the boundary conditions are fixed charge voltage and zero spin potentials defining the magnetically active device region. At the left reservoir, we have
\begin{equation}
V_l^{(c)}(z_l=0) = V/2 \,, \,\, \mathbf{V}_l^{(s)}(z_l=0)=0 
\end{equation}
and at the right reservoir, we have
\begin{equation}
\\V_r^{(c)}(z_r=L_r) = -V/2 \,, \,\, \mathbf{V}_r^{(s)}(z_r=L_r)=0 \, .
\end{equation}
The charge and spin accumulations are uniquely determined by Eqs.~(\ref{jc_1l})-(\ref{js_2r}), (\ref{diff_eq}), (\ref{diff_cur}), conservation of charges and spins close to the N$\mid$F interfaces in the normal metal, and the boundary conditions on the outer normal metal layers. 

We start from a parallel or antiparallel spin valve configuration in which magnetizations and spin accumulations are collinear to $\mathbf{x}$. In this initial configuration, the transport equations are easy to solve since both the spin accumulations and magnetizations directions are collinear throughout the whole circuit, and \textit{e.g.}\ the transverse components of the spin currents vanish. The current through the system then follows similarly to Ohm's law for resistances. 

We now consider small instabilities in the magnetization normal to the $x$ axis in Fourier space: $\mathbf{m}_{i}=\mathbf{x}+\delta \mathbf{m}_{i}^{\perp }(\mathbf{q})$. To linear order in the excitations, the charge current and voltage become $j^{(c)}=j^{(c0)}+\delta j^{(c)}(\mathbf{q})$ and $V^{(c)}=$ $V^{(c0)}+\delta V^{c}(\mathbf{q})$. We decompose the linearized spin potentials into longitudinal and transverse parts, $\delta \mathbf{V}^{(s)}=\mathbf{x}\delta V^{(s\parallel)}+\delta \mathbf{V}^{(s\perp )}$, expressing the spin transfer torque density $\mathbf{\tau}_{1}=\mathbf{j}_{1l}^{(s\perp)}-\mathbf{j}_{1m}^{(s\perp)}$ ($\mathbf{\tau}_{2}=\mathbf{j}_{2m}^{(s\perp)}-\mathbf{j}_{2r}^{(s\perp)}$) in terms of contributions from electrons hitting $F_{i}$ from the left and right: $\mathbf{\tau }_{1}=\mathbf{\tau }_{1l}+\mathbf{\tau }_{1m}$ ($\mathbf{\tau }_{2}=\mathbf{\tau }_{2m}+\mathbf{\tau }_{2r}$):
\begin{align}\text{\boldmath$\tau$}_{1l} & =2G_{1}^{\uparrow \downarrow }\left[V_{1l}^{(s0)}\delta \mathbf{m}_{1}^{\perp }-\delta \mathbf{V}_{1l}^{(s\perp)}\right] \, , \\\text{\boldmath$\tau$}_{1m} & =2G_{1}^{\uparrow \downarrow }\left[V_{1m}^{(s0)}\delta \mathbf{m}_{1}^{\perp }-\delta \mathbf{V}_{1m}^{(s\perp)}\right] \, , \\\text{\boldmath$\tau$}_{2m} & =2G_{2}^{\uparrow \downarrow }\left[V_{2m}^{(s0)}\delta \mathbf{m}_{2}^{\perp }-\delta \mathbf{V}_{2m}^{(s\perp)}\right] \, , \\\text{\boldmath$\tau$}_{2r} & =2G_{2}^{\uparrow \downarrow }\left[V_{2r}^{(s0)}\delta \mathbf{m}_{2}^{\perp }-\delta \mathbf{V}_{2r}^{(s\perp)}\right] \, .\end{align}
Solving the diffusion equations in each normal metal, we can also find the charge currents to linear order in the instability close to the N$\mid$F interfaces
\begin{align}
\delta j^{(c)}_l(L_l) & = G_l \frac{q L_l}{\tanh{qL_l}} \delta V_{1l}^{(c)} \, , \\
-\delta j^{(c)}_m(0)   & = G_m \frac{q L_m}{\tanh{qL_m}} \left[\delta V_{1m}^{(c)} - \frac{\delta V_{2m}^{(c)}}{\cosh{qL_m}} \right] \, , \\
\delta j^{(c)}_m(L_m) & = G_m  \frac{q L_m}{\tanh{qL_m}} \left[ \delta V_{2m}^{(c)} - \frac{\delta V_{1m}^{(c)}}{\cosh{qL_m}} \right] \, , \\
-\delta j^{(c)}_r(0) & = G_r \frac{q L_r}{\tanh{qL_r}} \delta V_{2r}^{(c)} \, , 
\end{align}
where $1/G_l=R_l=\rho_{N_l} L_l$, $1/G_m=R_m=\rho_{N_m} L_m$, and $1/G_r=R_r=\rho_{N_r} L_r$ are the resistance of the left, middle and right normal metal layers, respectively. The expressions for the spin-currents are similar by replacing $j^{(c)} \rightarrow \mathbf{j}^{(s)}$ and $V^{(c)} \rightarrow \mathbf{V}^{(s)}$.
The equations for the desired charge and spin potentials are can be obtained by noticing that expressions for $\delta j^{(c)}$ and $\delta j^{(s\parallel)}$ form a closed set of equations independent of $\delta \mathbf{m}_{\perp}$ with trivial solution: $\delta V^{(c)}=\delta V^{(s\parallel)}=0$ and $\delta j^{(c)}=\delta j^{(s\parallel)}=0$. Only transverse spin currents $\mathbf{j}^{(s\perp)}$ are therefore induced by small magnetization fluctuations. Finding the solutions of the linear equations for the unknown $\delta \mathbf{V}_{1l}^{(s\perp)}$, $\delta \mathbf{V}_{1m}^{(s\perp)}$, $\delta \mathbf{V}_{2m}^{(s\perp)}$, and $\delta \mathbf{V}_{2r}^{(s\perp)}$ is straightforward, but tedious. We present the results in the next section.

\section{Results and Discussion}\label{s:results}

In order to manage the complex analytical spin transfer torque expressions, we introduce the following standard notation for the $i$th ferromagnet: average resistance 
\begin{equation}
R_i^{\ast}=(R_i^{\uparrow }+R_i^{\downarrow })/4
\end{equation}
and polarization 
\begin{equation}
P_i=\frac{R_i^{\downarrow }-R_i^{\uparrow}}{ R_i^{\uparrow }+R_i^{\downarrow }} \, .
\end{equation}
The resistance contrast between parallel ($P_{1}P_{2}>0$) and anti-parallel ($P_{1}P_{2}<0$)configurations is 
\begin{equation}
\Delta R=R^{\text{ap}}-R^{\text{p}} = \frac{4R_{1}^{\ast }R_{2}^{\ast }\left\vert P_{1}P_{2}\right\vert }{(R_{l}+R_{1}^{\ast }+R_{m}+R_{2}^{\ast }+R_{r}} \, .
\end{equation} 
The polarization of the current is 
\begin{equation}
P_t = \frac{R_1^{\ast} P_1 + R_2^{\ast} P_2}{R_l + R_1^{\ast} + R_m + R_2^{\ast} + R_r} \, .
\end{equation}
The charge current is denoted $j^{(c)}$.
\subsection{Spin transfer torques}
The spin transfer torque on ferromagnet $1$ ($2$) has contributions from electrons hitting it from the left (middle) normal metal and from the middle (right) normal metal, $\text{\boldmath$\tau$}_{1} = \text{\boldmath$\tau$}_{1l} + \text{\boldmath$\tau$}_{1m}$ ($\text{\boldmath$\tau$}_{2} = \text{\boldmath$\tau$}_{2m} + \text{\boldmath$\tau$}_{2r}$). $\text{\boldmath$\tau$}_{1l}  =\mathcal{L}_{1l}\delta {\bf m}_{1}^{\perp }j^{(c)}$ on the first ferromagnet exerted by electrons coming from the left and $\text{\boldmath$\tau$}_{2r}=\mathcal{L}_{2r}\delta {\bf m}_{2}^{\perp }j^{(c)}$ on the second ferromagnet due to electrons from the right is given by:
\begin{equation}\mathcal{L}_{1l}=\frac{R_{l}\left[ f_{l}(0)-f_l(q)\right] }{R_{l}f_{l}(q)+R_{1}^{\uparrow\downarrow }} P_t  \label{L1l}.
\end{equation}
Expression for $\mathcal{L}_{2r}$ are obtained by changing sign and substituting $l$ by $r$ and $1$ by $2$. Here $f(x)=\tanh (x)/x$, $f_{l}(q)=f[x_{l}(q)]$, and $x_{l}(q)=(q^{2}+l_{l}^{-2})^{1/2}t_{Nl}$, where $t_{Nl}$ is the left normal metal thickness and $l_l$ its spin-diffusion length. In particular $f_l(0)\approx1$ (and similarly for $f_m$, $f_r$, and $x_m$, $x_r$). These torques acting on a single ferromagnetic layer vanish in the long wavelength limit as in Ref.~\onlinecite{Polianski:prl04}. The new physics is contained in the wave vector-dependent torques $\text{\boldmath$ \tau$}_{1m}$ and $\text{\boldmath$\tau$}_{2m}$. In the expressions $\text{\boldmath$\tau$}_{1m}=( \mathcal{L}_{1m}\delta {\bf m}_{1}^{\perp }+\mathcal{K}_{1m}\delta {\bf m}_{2}^{\perp }) j^{(c)}$ and $\text{\boldmath$\tau$}_{2m}=(\mathcal{L}_{2m}\delta {\bf m}_{2}^{\perp}+\mathcal{K}_{2m}\delta {\bf m}_{1}^{\perp }) j^{(c)}$, the spin transfer torques due to spin currents between the two ferromagnets are found as: 
\begin{equation}\mathcal{L}_{1m}=\frac{-\left(n_{1}+n_{2}x_{m}^{2}+n_{3}/f_{m}\right) /R^{\ast }}{R_{m}^{2}+R_{1}^{\uparrow \downarrow }R_{2}^{\uparrow \downarrow}x_{m}^{2}+R_{m}\left( R_{1}^{\uparrow \downarrow }+R_{2}^{\uparrow\downarrow }\right)/f_{m}}\,.  \label{L1m} \end{equation}
where, 
\begin{align}
n_{1} & =  -R_{m}^{2}( P_{1}R_{1}^{\ast }+P_{2}R_{2}^{\ast}) \, , \\
n_{2} & =  R_{2}^{\uparrow \downarrow }[ P_{1}R_{1}^{\ast}( R_{2}^{\ast }+R_{m}+R_{r}) -P_{2}R_{2}^{\ast }(R_{l}+R_{1}^{\ast })] \, ,  \\
n_{3} & = n_1 R_2^{\uparrow \downarrow}/R_m + n_2 R_m/R_2^{\uparrow \downarrow} 
\end{align}
introducing $R^{\ast }=R_{l}+R_{1}^{\ast }+R_{m}+R_{2}^{\ast }+R_{r}$. The expressions for $\mathcal{L}_{2m}$ is similar to that for $\mathcal{L}_{1m}$, with an overall sign change and the substitution $l \leftrightarrow r$ and $1 \leftrightarrow 2$. $\mathcal{K}_{1m}$ and $\mathcal{K}_{2m}$ govern the dynamic coupling between the ferromagnets, but do not affect the instabilities of individual ferromagnets and are therefore not discussed here. In the limit of long-wavelength excitations, $q\to0$, our results reduce to previous ones.\cite{Brataas:prl00,Manschot:apl04}

Let us discuss the spin transfer torques (\ref{L1l}) and (\ref{L1m}) in simple limits. 

(i) When the ferromagnet 2 and the right normal metal are absent in the circuit, \textit{e.g.} when $R_2^{\ast}=0$, $R_2^{\uparrow \downarrow}=0$, and $R_r=0$,  we naturally obtain a similar result as the spin-torque on the first ferromagnet exerted by electrons coming from the left (\ref{L1l})\cite{Polianski:prl04}
\begin{equation}
{\cal L}_{1m} = - \frac{R_m \left[f_m(0) - f_m(q) \right]}{R_m f_m(q) + R_1^{\uparrow \downarrow}} P_t \, .
\end{equation}
When, additionally, the system is symmetric around ferromagnet 1, $R_l=R_m$, the total spin transfer torque on ferromagnet 1 vanishes, $\mathcal{L}_{1l}+ \mathcal{L}_{1m}=0$.\cite{Polianski:prl04}

(ii) For \textit{symmetric} junctions in a parallel magnetic configuration $R_{l}=R_{r}$, $R_{1}^{\uparrow \downarrow }=R_{2}^{\uparrow \downarrow }$, $R_{1}^{\ast }=R_{2}^{\ast }$, and $P_{1}=P_{2}$. For macrospin ($q=0$) excitations 
\begin{align}
\mathcal{L}_{1l} & =0 \, , \\
\mathcal{L}_{1m} & =P_t/2 \, , 
\end{align}
illustrating that the spin transfer torque is exerted by electrons coming from ferromagnet 2 that hit ferromagnet 1 from the right and is governed by the polarization of the entire spin-coherent region $P_t$ (Ref. \onlinecite{Manschot:apl04}). At short wavelengths, 
\begin{align}
\mathcal{L}_{1l} & =P_t \frac{R_{l}}{R_{1}^{\uparrow \downarrow}} \, , \\
\mathcal{L}_{1m} & =- P_t \frac{R_{m}}{2R_{1}^{\uparrow \downarrow}} \, , 
\end{align} 
so that 
\begin{equation}
\mathcal{L}_{1}=\mathcal{L}_{1l}+\mathcal{L}_{1m}=-P_t \frac{R_{m}-2R_l}{2 R_{1}^{\uparrow \downarrow}} \, .
\end{equation}
For the dynamics of a single ferromagnet, it is the asymmetry in the diffusion process to the left and to the right of the ferromagnet that determines the sign of the short wavelength spin transfer torque.\cite{Polianski:prl04} We see that for spin-valves, it is the asymmetry of the complete device that determines when short wavelength modes cannot be excited. When $R_m=2R_l$, small wavelength modes cannot be excited. This condition ($R_m=2R_l$) corresponds to a completely symmetric spin valve, where two identical effective "ferromagnets" consists of e.g. $R_l$, ferromagnet 1 and half of ferromagnet $R_m$ and half of ferromagnet $R_m$, ferromagnet 2 and $R_r$. In the limit $R_l=0$ there is no short wavelength spin transfer torque from electrons that hit ferromagnet 1 from the left and vice versa when $R_m=0$ we find $\text{\boldmath$ \tau$}_{1m}( q\rightarrow 0) \rightarrow 0$.  On the other hand, the macrospin, torque in symmetric junctions only depends on the global polarization for symmetric junctions.\cite{Manschot:apl04}. We thus provided proof of the conjecture in Ref.~\onlinecite{Ozyilmaz:04} that the spin transfer torque $\mathcal{L}_{1m}$ can change its sign as a function of $q$ when $R_m > 2R_l$ for symmetric spin-valves. In that case, the magnetization moves as a macrospin for one current direction but short wavelength spin waves are excited when the current is reversed. In general, we find that when the normal-metal bulk resistance asymmetry is smaller than the interface spin-mixing resistance the macrospin spin transfer torque is larger than the short wavelength spin transfer torque. For strongly asymmetric structures, $\mathcal{L}_{1m}$ is negative also for long-wavelength excitations.\cite{Manschot:apl04}

\subsection{Enhanced Gilbert damping}
The critical current for the onset of magnetic instabilities depends also on spin-pumping by a moving magnetization that opposes the dynamics.\cite{Tserkovnyak:prl02} Spin pumping by ferromagnet $i$ into neighboring normal metals $\propto G_{i}^{\uparrow \downarrow } \mathbf{m}_{i}\times \frac{\partial \mathbf{m}_{i}}{\partial t} $ enhances the Gilbert damping $\alpha _{1}=\alpha_{1}^{(0)}+\alpha _{1l}^{\prime}+\alpha _{1m}^{\prime}$ and $\alpha _{2}=\alpha_{2}^{(0)}+\alpha _{2m}^{\prime}+\alpha _{2r}^{\prime}$, where $\alpha_1^{(0)}$ and $\alpha_2^{(0)}$ are the damping parameters in the isolated ferromagnets. Spin pumping via the left normal metal to the left reservoir gives \cite{Polianski:prl04} 
\begin{equation}
\alpha _{1l}^{\prime}= \frac{\gamma^*}{8 \pi M_1 V_1} \frac{ R_{K}}{ R_{1}^{\uparrow \downarrow }+R_{l}f_{l}(q)} \,  ,
\end{equation}
where $\gamma ^*$ is the gyromagnetic ratio, $M_{i}$ and $V_{i}$ are the magnetization and volume of ferromagnet $i$,  and $R_{K}=h/e^{2}$ is a conductance quantum. The enhancement due to spin pumping via the middle normal metal to ferromagnet 2 comprises a new result:
\begin{equation}\alpha _{1m}^{\prime}=  \frac{\eta_1R_{K}\left( R_{2}^{\uparrow \downarrow }x_{m}^{2}+R_{m}/f_{m}\right) }{R_{m}^{2}+R_{1}^{\uparrow \downarrow }R_{2}^{\uparrow \downarrow}x_{m}^{2}+R_{m}\left( R_{1}^{\uparrow \downarrow }+R_{2}^{\uparrow\downarrow }\right)/f_{m}} \, .\label{a1m}
\end{equation}
Both $\alpha_{1l}^\prime$ and $\alpha_{1m}^\prime$ can be understood in terms of an effective resistance against pumping spins out of the ferromagnet. Spin pumping to the left is limited by the wavelength-dependent effective total resistance $R_{1}^{\uparrow \downarrow }+R_{l}f_{l}$. In the long-wavelength limit the resistors are in series, $R_{1}^{\uparrow \downarrow }+R_{l}$. In the short-wavelength limit, the effective resistance is reduced due to the inhomogeneous spin distribution. Spin pumping to the right, in the long-wavelength limit governed by $R_{1}^{\uparrow \downarrow }+R_{m}+R_{2}^{\uparrow\downarrow }$ is also reduced in the short-wavelength limit. For the second ferromagnet, we compute $\alpha_{2}=\alpha _{2}^{(0)}+\alpha _{2l}^{\prime}+\alpha _{2r}^{\prime}$ with identical expression as above with $1 \leftrightarrow 2$ and $l \leftrightarrow r$. 
\subsection{Critical Currents for magnetic excitations}
Including spin transfer torques and enhanced damping, the magnetization dynamics obey a generalized Landau-Lifshitz-Gilbert equation:
\begin{equation}\frac{\partial {\mathbf m}_{i}}{\partial t}=-\gamma {\mathbf m}_{i}\times {\bf H}_{\text{eff}}+\frac{\gamma ^*}{2eM_{i}t_{Fi}} {\boldmath \text{$\tau $}}_{i}+\alpha _{i} {\mathbf m}_{i}\times \frac{\partial {\mathbf m}_{i}}{\partial t}.
\end{equation}
The effective magnetic field ${\mathbf H}_{\text{eff}}$ is the functional derivative of the total magnetic energy $U$ due to external magnetic field, magnetic anisotropy and spin-wave stiffness. To lowest order in the excitation amplitude $U=-{\mathbf m}_{i}^{(0)}\cdot {\bf x} H+K_{1}({\bf u}_{1}\cdot \delta {\mathbf m}_{i}^{\perp })^{2}+K_{2}({\bf u}_{2}\cdot \delta {\mathbf m}_{i}^{\perp })^{2}$, where the applied magnetic field $H$ is assumed to be along the initial magnetization direction. $K_{1}(q)$ and $K_{2}(q)$ are wave number dependent spin-wave stiffnesses that for $q=0$ reduce to the anisotropy constants along axes $\mathbf{u}_1$ and $\mathbf{u}_2$. A configuration in which ferromagnet 1 is aligned parallel with the external magnetic field becomes unstable when $j^{(c)}>j_{1}^{+}$, whereas an antiparallel magnetization becomes unstable when $j^{(c)}<j_{1}^{-}$. We compute the critical currents at $T=0$ requiring that the damping torque is exactly canceled by the spin transfer torque:
\begin{equation}j_{1}^{\pm }=\frac{\alpha _{1}^{(0)}+\alpha _{1l}^{\prime}+\alpha_{1m}^{\prime}}{\mathcal{L}_{1l}^{\pm }+\mathcal{L}_{1m}^{\pm }}\frac{2e}{\hbar }M_{1}t_{F1} \frac{\omega^{\pm }}{\gamma ^*} \, ,
\end{equation}
introducing the ferromagnetic resonance frequencies $\omega^{\pm}=H\pm (K_{1}+K_{2})/2$. $\mathcal{L}_{1l}^{\pm }$ and $\mathcal{L}_{1m}^{\pm}$ denote spin transfer torques computed for positive/negative polarization, $P_{1}= \pm \left\vert P_{1}\right\vert $.  When the torques change sign as a function of wave vector the ferromagnet can be unstable against macrospin excitations when the current is flowing in one direction, but spin-waves are excited when the current is reversed, as pointed out above. Similar expressions hold for $j_2^{\pm}$ the critical current for ferromagnet 2, $j_2^{\pm}$.
\subsection{Comparison with NYU/IBM experiments}
We are now in position to discuss the experiments in Ref. \onlinecite{Ozyilmaz:04}. A good measure of the efficiency of the spin transfer torques is the critical current $\partial j_i/\partial H$ at high magnetic fields when anisotropies $K_1$ and $K_2$ do not play a role.\cite{Sun:mmm04} The magnetoresistance, enhanced Gilbert damping, spin transfer torques and switching current all depend strongly on the detailed device parameters. We adopt resistance parameters collected by the Michigan State University group for Co/Cu systems,\cite{Bass:mmm99} \textit{i.e.}, $\varrho _{\text{Co}}^{\ast }=75$~n$\Omega $m, bulk polarization $\beta=0.46$, $\varrho_{\text{Cu}}^{\ast }=6$~n$\Omega $m, $R_{\text{Co/Cu}}^{\ast}=0.51$~f$\Omega $m$^{-2}$, and an interface polarization $\gamma =0.77$. The mixing conductance that agree with first-principles band-structures calculations and ferromagnetic resonance experiments is $R_{\text{Co/Cu}}^{\uparrow \downarrow }=1.0$~f$\Omega $m$^{-2}$ (Ref. \onlinecite{Tserkovnyak:prl02}). The magnetically active region is 10Cu$\mid$3Co$\mid$10Cu$\mid$12Co$\mid$35 Cu, where the numbers denote the thickness of the layers in nanometers. The right normal metal thickness is chosen smaller than the geometrical since the nano-pillar widens.\cite{Ozyilmaz:04}

\begin{figure}[tbp]
\includegraphics[width=\linewidth,angle=0,clip=true,width=8cm]{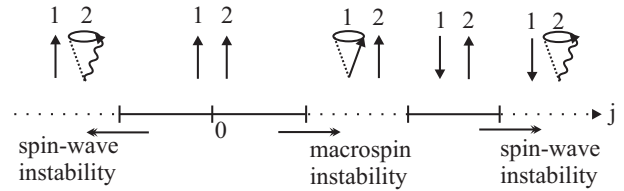} \caption{Phase diagram for ferromagnet 1 and 2, starting in a parallel configuration, as a function of current.}
\label{phase}
\end{figure}

\begin{figure}[tbp]
\includegraphics[width=\linewidth,angle=0,clip=true,width=8cm]{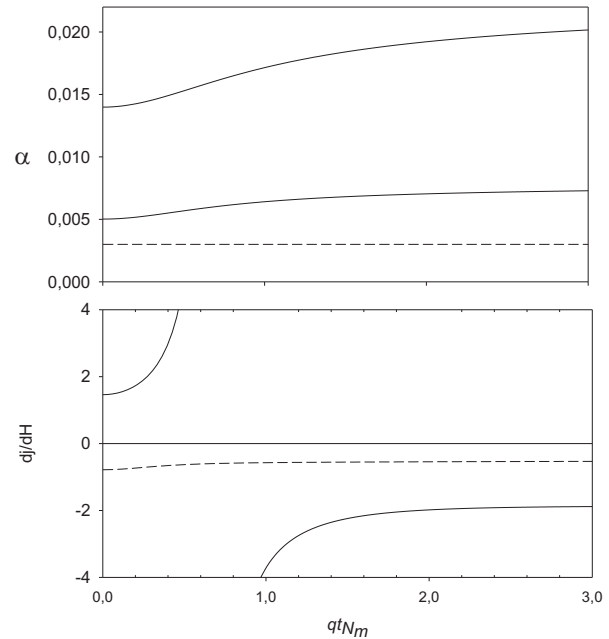}
\caption{Upper: Gilbert damping in thin ferromagnet 1 (upper curve), thick ferromagnet 2, and in bulk (dashed line) vs. $qt_{Nm}$. Lower: Critical current $\partial j/\partial H$ in units of 10$^{12}$Am$^{-2}$T$^{-1}$ of thin ferromagnet 1 (solid line) and thick ferromagnet 2 (dashed line) vs. $qt_{Nm}$.}
\label{alpha}
\end{figure}

Before going into details, we summarize our results for the NYU/IBM samples in Fig.\ \ref{phase} in semi-quantitative agreement with the experiments. Starting in a parallel configuration there is a macrospin instability for ferromagnet 1 at positive currents. The macrospin excitation leads to an antiparallel configuration for larger currents,\cite{Ozyilmaz:prl03} which should be confirmed by spatially dependent micromagnetic calculations. A further increase in the current leads to a spin-wave instability in ferromagnet 2. Similarly, for negative current, we predict a spin-wave instability for ferromagnet 2 (Ref. \onlinecite{Kiselev:new}).
The theoretical Gilbert damping constant is plotted in Fig.~\ref{alpha} for a bulk Gilbert damping $\alpha _{\text{Co}} ^{(0)}=0.003$ for both ferromagnets. Spin-pumping is important, giving rise to a strongly enhanced Gilbert damping constant of the thin ferromagnet that increases with wave vector. Both spin-pumping and spin transfer torque increase with the interface to volume ratio. Ferromagnet 1 is easier to excite but its enhanced damping partially compensates this effect, thus allowing excitations of both ferromagnets to compete.
Numerical results for the critical current $\partial j_{1}/\partial H$ at high magnetic fields are displayed in Fig. \ref{alpha} for the thin ferromagnet 1 and the thick ferromagnet 2 for the parallel configuration. For positive currents, experiments find switching of the thin ferromagnet presumably to an antiparallel configuration when $\partial j^{(c)}/\partial H=0.6\times 10^{12}$~Am$^{-2}$T$^{-1}$ (Ref. \onlinecite{Ozyilmaz:04}). This value is extracted from the line in their contour plot of $d^2V/d^2I$ showing the onset of magnetic excitations. Indeed, for positive currents macrospin excitations occur first with lowest critical current that agrees well with experiments. Furthermore, we expect spin-wave excitations for the thick ferromagnet 2 at opposite currents, again in good agreement with experiments.\cite{Ozyilmaz:04} Additionally (but not shown here) we predict that after the thin ferromagnet 1 switches, ferromagnet 2 also becomes unstable against spin-wave excitations for positive currents, supported by experiments as well. 

Residual quantitative discrepancies between experiments and theory are believed to be caused by uncertainties in the material and devices parameter and details in the micromagnetic structure. We have used measurements for \textit{e.g.}\ the bulk resistivities from the MSU group and it is likely that due to a different sample preparation technique, the NYU/IBM samples have different resistivities. Furthermore, the thickest ferromagnet is 12nm in the NY/IBM measurements, which is at the boundary where spin-wave excitations along the transport direction become of importance in this layer, which reduces the accuracy of the predicted value of the critical current in the thick layer.\cite{Stiles:prb04} Given these uncertainties, we consider the semi-quantitative agreement encouraging.

\section{Conclusions}\label{s:conclusions}
In conclusion, we report a theory of macrospin \textit{vs.} spin-wave excitations in spin valves that explains recent observations using only independently determined material and device parameters. The rich phase space of magnetic excitations is classified in terms of macrospin and finite wavelength spin-wave excitations that depend on the resistance distribution in the magnetically active region. For symmetric junctions, macrospin instabilities are strongly favored. Finite wavelength spin-wave excitations are pronounced in asymmetric spin-valves with relatively high normal metal resistances comparable to that of the ferromagnets. Since the results are in agreement with recent experiments, we are confident that our insights should be helpful to explore the magnetization dynamics in the full parameter space spanned by currents, external magnetic fields, and device design.

\acknowledgments 
A.\ B.\ thanks A.\ D.\ Kent, B.\ \"{O}zyilmaz, W. Chen, M. Stiles and J.\ Z.\ Sun for stimulating discussions. This work has been supported in part by the Research Council of Norway, NANOMAT Grants. No 158518/143 and 158547/431, the Harvard Society of Fellows, the FOM, and EU via NMP2-CT-2003-505587 'SFINX'.


\begin{thebibliography}{99}
\bibitem{Berger:prb96} L. Berger, Phys. Rev. B {\bf 54}, 9353 (1996). 
\bibitem{Slonczewski:mmm96} J. Slonczewski, J. Magn. Magn. Mater. {\bf 159}, L1, (1996).
\bibitem{Tsoi:prl98} M. Tsoi \textit{et al.}, Phys. Rev. Lett. \textbf{80}, 4281 (1998); J.-E. Wegrowe \textit{et al.}, Europhys. Lett. \textbf{45}, 626(1999); E.B. Myers \textit{et al.}, Science \textbf{285}, 867(1999); J.A. Katine \textit{et al.}, Phys. Rev. Lett. \textbf{84}, 3149 (2000); J. Grollier \textit{et al.},Appl. Phys. Lett. \textbf{78}, 3663 (2001); S. Urazhdin \textit{et al.} Phys. Rev. Lett. \textbf{91}, 146803 (2003); S. I. Kiselev \textit{et al.} Nature (London) \textbf{425}, 380 (2003); I. N. Krivotorov \textit{et al.} Science \textbf{307}, 228 (2005).
\bibitem{Ozyilmaz:prl03} B. \"{O}zyilmaz \textit{et al.} Phys. Rev. Lett. \textbf{91}, 067203 (2003).
\bibitem{Brataas:prl00} A. Brataas, Yu.V. Nazarov, and G.E.W. Bauer, Phys. Rev. Lett. \textbf{84}, 2481 (2000); Eur. Phys. J. B \textbf{22}, 99 (2001); K. Xia \textit{et al.}, Phys. Rev. B \text{65}, 220401 (R) (2002); G. E. W. Bauer \textit{et al.}, Phys. Rev. B \textbf{67}, 094421 (2003);  M. Zwierzycki \textit{et al.}, cond-mat/0402088.
\bibitem{Waintal:prb00} X. Waintal \textit{et al.}, Phys. Rev. B \textbf{62}, 12317 (2000); M.~D. Stiles and A. Zangwill, Phys. Rev. B \textbf{66}, 014407 (2002).
\bibitem{Wernsdorfer} W. Wernsdorfer \textit{et al.} Phys. Rev. Lett. \textbf{78}, 1791 (1997); R. H. Koch \textit{et al.} Phys. Rev. Lett. \textbf{84}, 5419 (2000).


\bibitem{Tserkovnyak:prl02} Y. Tserkovnyak, A. Brataas, and G.~E.~W. Bauer, Phys. Rev. Lett. \textbf{88}, 117601 (2002); Phys. Rev. B \textbf{67}, 140404(R) (2003); for a review, see cond-mat/040924.
\bibitem{Apalkov} D. M. Apalkov and P. B. Visscher, cond-mat/0405305; T. Valet, unpublished; G. Bertotti \textit{et al.}, Phys. Rev. Lett. \textbf{94}, 127206 (2005).
\bibitem{Polianski:prl04} M. L. Polianski and P. W. Brouwer, Phys. Rev. Lett. \textbf{92}, 026602 (2004); S. Adam, M. L. Polianski, and P. W. Brouwer, cond-mat/0508732.
\bibitem{Stiles:prb04} M. D. Stiles, J. Xiao, and A. Zangwill, Phys. Rev. B \textbf{69}, 054408 (2004); T.Y. Chen \textit{et al.} Phys. Rev. Lett \textbf{93}, 026601 (2004); B. Ozyilmaz \textit{et al.}, Phys. Rev. Lett \textbf{93} 176604 (2004).
\bibitem{Ozyilmaz:04} B. \"{O}zyilmaz \textit{et al.} Phys. Rev. B \textbf{71}, 140403(R) (2005).
\bibitem{Lee:natmat04} K.J. Lee \textit{et al.}, Nature Materials \textbf{3}, 877(2004); Z. Li and S. Zhang, Phys. Rev. B \textbf{68}, 024404 (2003).

\bibitem{Manschot:apl04} J. Manschot, A. Brataas, and G. E. W. Bauer, Appl. Phys. Lett. \textbf{85}, 3250 (2004).
\bibitem{Bass:mmm99} J. Bass and W. P. Pratt Jr., J. Magn. Magn. Mater. \textbf{200}, 274 (1999).
\bibitem{Sun:mmm04} J. Z. Sun et al., proceedings of the 49th MMM 2004.

\bibitem{Kiselev:new} After submitting our orignal article cond-mat/0501672, the Cornell collaboration, S. I. Kiselev et al., cond-mat/0504402 v.1 and v.2, published high-frequency measurements that provide direct evidence that the interpretation in Ref.\ \onlinecite{Ozyilmaz:04} and the present calculations are correct. W. Chen et al, cond-mat/0509034, subsequently confirm that bipolar excitations are present only in asymmetric spin valves, in agreement with our theory.

\end{thebibliography}
\end{document}